\documentclass[twocolumn,amsmath,aps]{revtex4}
\usepackage{graphicx}

\newcommand{\nl}{\nonumber \\}

\newcommand{\be}{\begin{equation}}
\newcommand{\ee}{\end{equation}}
\newcommand{\bea}{\begin{eqnarray}}
\newcommand{\eea}{\end{eqnarray}}

\newcommand{\Eq}[1]{Eq.\,(\ref{#1})}
\newcommand{\Eqs}[1]{Eqs.\,(\ref{#1})}

\newcommand{\la}{\langle}
\newcommand{\ra}{\rangle}
\newcommand{\dg}{\dagger}

\newcommand{\ti}{\tilde}
\newcommand{\mb}{\mbox}

\begin{document}
\draft

\title{ Measurement of single electron spin with sub-micron Hall magnetometer }

\author{ Jinshuang Jin and Xin-Qi Li}
\address{Institute of Semiconductors,
         Chinese Academy of Sciences, P.O.~Box 912, Beijing 100083, China}

\date{\today; Ms.\# L04-5566}

\begin{abstract}
Sub-micron Hall magnetometry has been demonstrated as an efficient
technique to probe extremely weak magnetic fields.
In this letter, we analyze the possibility of employing it
to detect single electron spin. Signal strength and readout time
are estimated and discussed with respect to a number of practical issues.
\end{abstract}
\maketitle

Hall probes have been employed in studies of magnetic properties of
materials for several decades.
Based on the transport properties of quasi-one-dimensional (Q1D)
Hall bar system under non-uniform magnetic field \cite{Li98},
the Hall technique has been extended to sub-micron probes for
{\it individual} microfabricated samples \cite{Gei98}.
It was found in several regimes the Hall magnetometer has
advantages over the alternative techniques such as
$\mu$-SQUIDs \cite{Cha91} and $\mu$-mechanical cantilevers \cite{Sid95}.


In recent years, largely being stimulated by the interest of
solid-state quantum computation, measurement of single electron spin
is becoming an intensive research subject.
In particular,
it has been suggested that the technique based on the magnetic resonance
force microscopy (MRFM), i.e., the $\mu$-mechanical cantilever technique,
is probably one of the most promising means \cite{Rug0103}.
In concern with the sensitivity of the micro-Hall magnetometer, it
has allowed to register magnetic changes of
$10^{3}\sim 10^{5}~\mu_{\mb{\tiny B}}$ in the sensitive area of cross
junction ($\mu_{\mb{\tiny B}}$ is the Bohr magneton)\cite{Gei98},
which is similar to the demonstrated sensitivity of MRFM technique.
Remarkably, recent progress on the MRFM is seemingly to allow the possibility
to register single electron spin \cite{Rug04}.
In this letter, we address the issue to detect single electron spin
based on the sub-micron Hall magnetometer.

The measurement setup is shown schematically in Fig.\ 1,
where the magnetic dipole (electron spin) to be measured
is placed at $z_0$ above the cross center of the Hall magnetometer.
Due to the presence of the magnetic dipole, non-zero magnetic field
exists in the cross region of the Hall junction.
The Hall-meter senses this magnetic field by relating it with
the Hall signal (Hall resistance or Hall voltage).
Conceptually, the Hall signal depends on the dipole (spin-up or
spin-down) state, which in physical principle is a counterpart of the
well-studied problem of charge qubit (quantum bit) measured
by quantum-point-contact (QPC)\cite{Gur97} or single-electron-transistor
(SET) \cite{Sch98}.

For simplicity, we assume four identical leads which are fabricated
from the two-dimensional electron gas (2DEG),
each having a width $W$ and connecting to an electron
reservoir with chemical potential $\mu_i$. In the linear response
regime, from the Landauer-B\"uttiker formula \cite{But86},
the Hall resistance can be expressed in terms of the
transmission probabilities $T_{ij}$ (from lead ``$j$" to lead ``$i$")
\bea\label{RH-1}
R_H=\frac{(\mu_2-\mu_4)/e}{I}=\frac{h}{2e^2}
    \frac{T_{21}-T_{41}}{Z}  ,
\eea
where $Z=T^2_{21}+T^2_{41}+2T_{31}(T_{31}+T_{21}+T_{41})$.
Here, the current boundary condition $I_1=-I_3=I$ and $I_2=I_4=0$
are adopted for the Hall measurement.
\begin{figure}\label{Fig1}
\begin{center}
\centerline{\includegraphics [scale=0.35]{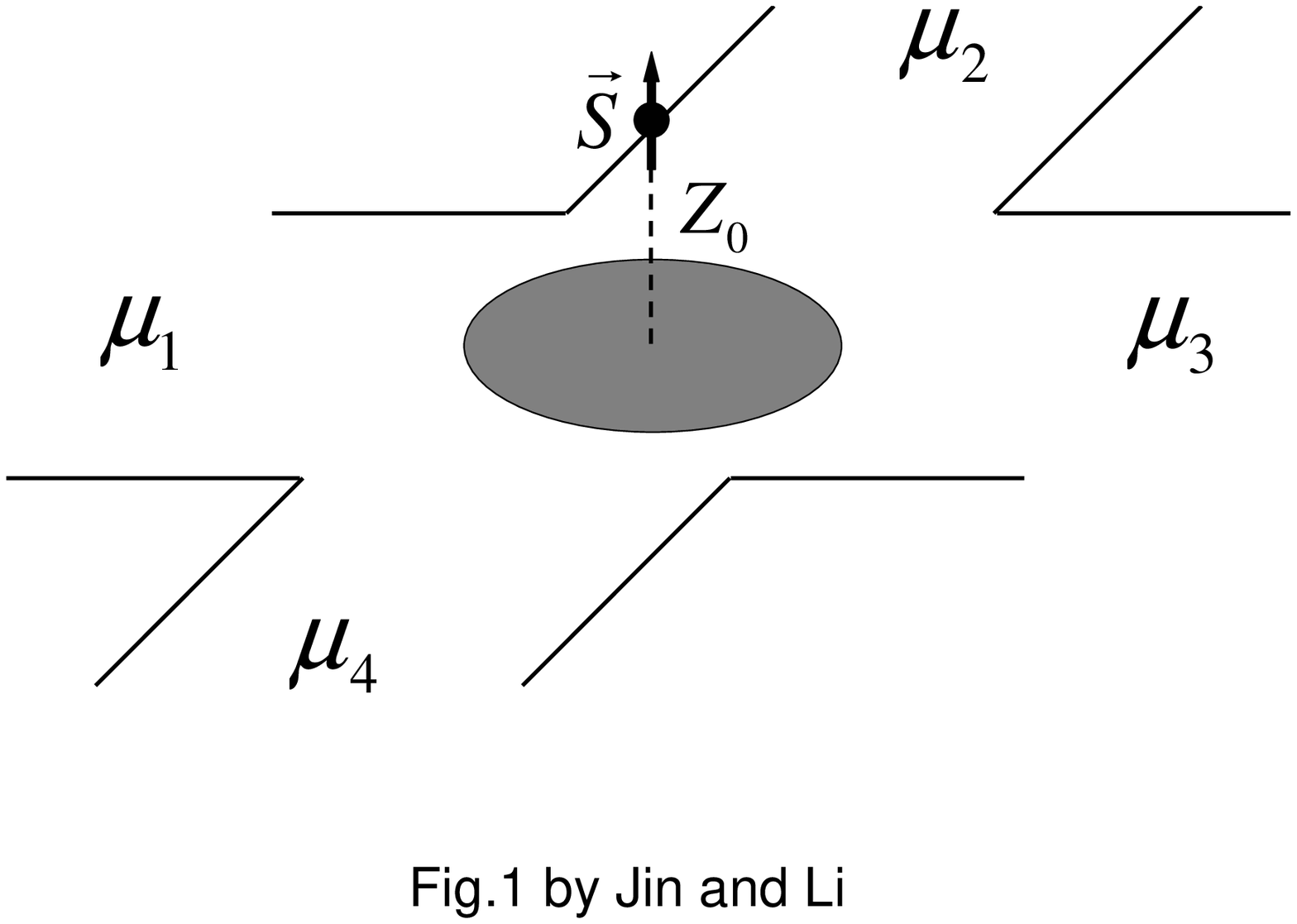}} \caption{
Schematic setup for the sub-micron Hall magnetometer employed to
detect single electron spin. }
\end{center}
\end{figure}

In this work, we focus on the extremely low temperature
which is required for any quantum measurement.
In this regime, the electron motion in the Hall bar is ballistic,
and a semi-classical Monte Carlo simulation is applicable
to compute the transmission probabilities $T_{ij}$ \cite{Li98}.
This approach can easily handle arbitrary inhomogeneous magnetic-field
profiles in the Hall cross.
It was found that in the regime of low magnetic field, the Hall resistance
is determined by the average magnetic field in the cross junction,
and is independent of the shape and position of the profile in the junction.
In particular, in the weak magnetic field regime
a {\it universal relation} was found between the Hall resistance
and the average magnetic field $\la B\ra$ in the junction region \cite{Li98}:
\bea\label{RH-2}
R_H=\frac{\la B\ra}{2}\frac{R_0}{B_0} ,
\eea
where $R_0=(h/2e^2) \pi/k_F W$, and $B_0=mv_F/eW$.
The Fermi wave-vector and velocity is related with the Fermi energy via
$E_F=\hbar^2k^2_F/2m=mv^2_F/2$, where $m$ is the effective electron mass
of the 2DEG.
\begin{figure}[h]
\begin{center}
\centerline{\includegraphics [scale=0.4]{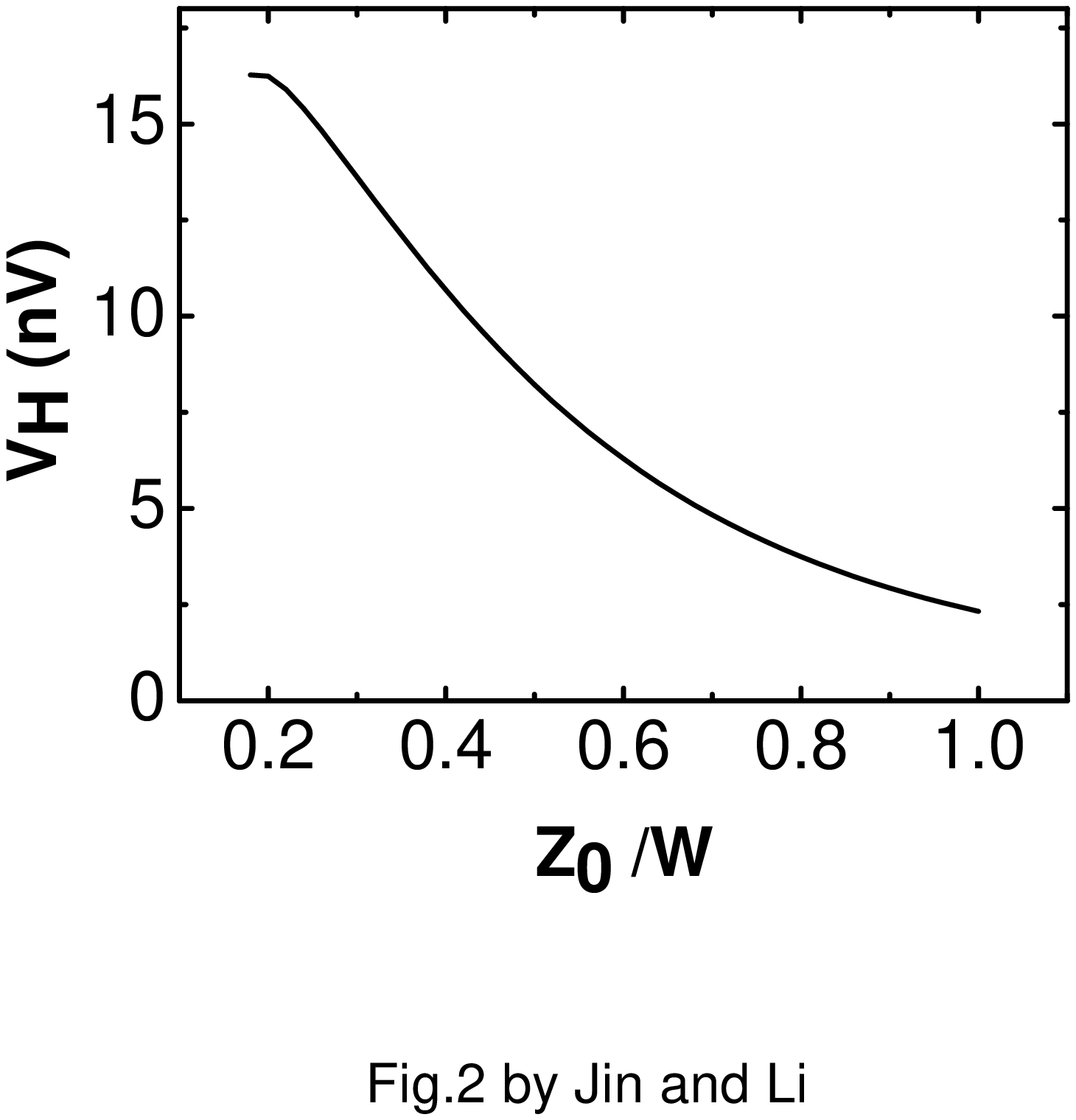}}
\end{center}
\caption{ Hall voltage versus the distance $z_0$ where the electron spin
is placed above the cross center. Other parameters are referred to
the main text. }
\end{figure}

Combining \Eqs{RH-1} and (\ref{RH-2}), we are able to relate the Hall voltage
with the parameters of the Hall magnetometer.
First, we obtain
\bea\label{dT}
T_{21}-T_{41}=\frac{Z\pi\hbar e}{4m}
   \frac{\la B\ra}{E_F}\equiv C_1\frac{\la B\ra}{E_F}.
\eea
In the regime of (extremely) weak magnetic field as considered in this letter,
``$Z$" can be well approximated to be a constant independent of the
magnetic field, thus ``$C_1$" is a constant.
On the other hand, based on the Landauer-B\"uttiker formula
\cite{But86}, it is straightforward to derive
$V_H=[(T_{21}-T_{41})/C_2] V_{13}$, where $V_{13}=(\mu_1-\mu_3)/e$
is the voltage applied across the leads ``1" and ``3", and
$C_2\equiv T_{21}+T_{41}+2T_{31}$ is also a magnetic-field
independent constant in the weak magnetic field regime.
\vspace{5ex}
Accordingly, we arrive at a simple expression for the Hall voltage
\bea\label{VH}
V_H=\frac{C_1}{C_2}\frac{\la B\ra}{E_F} V_{13} .
\eea
This result clearly shows that the Hall signal (voltage) is determined
by the average magnetic field $\la B\ra$ and the Fermi energy $E_F$ of
the confined 2DEG.
Note that $\la B\ra=\Phi/W^2$, where $\Phi$ is the magnetic flux
penetrating through the cross region.
For the single electron spin measured by the Hall magnetometer,
the magnetic flux is distributed in a limited range of nanometers
in the Hall cross.
We then suggest that the Hall junction
should be designed with narrow width ``$W$" in order to enhance the
Hall signal. Also, for the same purpose, small Fermi energy $E_F$
is favorable, which can be achieved by doping relatively low density
of electron numbers as the 2DEG.

To carry out a quantitative estimate, we consider the 2DEG formed from
the GaAs heterostructure, which has an effective electron mass
of $m=0.067m_e$.
Other parameters assumed are the equilibrium Fermi energy $E_F=10$ meV
which corresponds to electron-number density of
$n_e=2.8\times 10^{11}$ cm$^{-2}$,
and the junction channel width $W=10$ nm.
To estimate the average magnetic field in the cross region,
consider $z_0=5$ nm at which the electron is located above
the cross center, giving rise to
$\langle B\rangle\simeq 8.5\times 10^{-2}G$.
Using these parameters, we can easily compute the four relevant
transmission probabilities $T_{j1}$ ($j=1,\dots,4$),
based on \Eq{dT} and
(i) the probability conservation condition $\sum_j T_{j1}=1$,
and (ii) the values of $T_{11}\simeq 2.5\times 10^{-4}$
and $T_{31}\simeq 0.41$,
which are obtained from direct numerical simulations.
We thus arrive at an estimate for the Hall voltage,
$V_H\simeq 10^2 eV_{13}/E_F$ (in units of nano-volts).
As an example, let us set the ratio $eV_{13}/E_F=0.1$,
which leads to a Hall voltage of $\sim 10$ nV.
The result for a wider range of parameter is shown
in Fig.\ 2, where the dependence of $V_H$ on $z_0/W$ is plotted
(with other parameters unchanged as above).
We notice that the Hall signal about 10 nV will be within the access of
modern technology of voltage measurement, provided other noisy fluctuations
can be well suppressed.

In addition to the above estimate of signal strength,
below we address the issue of quantum measurement,
by regarding the spin as a quantum bit (qubit).
Suppose that the spin is subject to an operation
(i.e. undergoing Rabi oscillation).
We then consider the influence of the Hall measurement on it.
The spin Hamiltonian reads $H_{S}=\epsilon_\uparrow
|\uparrow\ra\la \uparrow| + \epsilon_\downarrow |\downarrow\ra\la
\downarrow| + \Omega(|\downarrow\ra\la \uparrow|+|\uparrow\ra\la
\downarrow|) $.
Correspondingly, the electron transport in the Hall magnetometer
is governed by $H_D=H_R+H_T$.
Here $H_R=\sum_{m=1} ^{4} \sum_k\epsilon_{mk} c^{\dg}_{mk}c_{mk}$
is the Hamiltonian of the four
electron reservoirs, and the tunneling Hamiltonian reads
$H_T=\sum_{m,n}(t_{mn}|\downarrow\ra\la \downarrow|
+\ti{t}_{mn}|\uparrow\ra\la \uparrow|)
\left[\sum_{kq}(c^{\dg}_{mk}c_{nq}+\rm{H.c.})\right]$,
where the summation is over all the transport channels
``$(m,n)=(1,2),(1,3),(1,4),(2,3),(2,4),(3,4)$".
Alternatively, let us re-express $H_T$ as
$H_T=\sum_{\alpha=\uparrow,\downarrow}Q_{\alpha}F_{\alpha}$,
where $Q_{\uparrow}=|\uparrow\rangle\langle\uparrow|$,
$Q_{\downarrow}=|\downarrow\rangle\langle\downarrow|$,
and the corresponding $F_{\alpha}$ can be accordingly determined.
In this form, the measurement device is clearly playing a role
of dissipative environment, and the measured spin would suffer
dephasing and relaxation owing to the back-action of measurement.
Precisely following Ref.\ \onlinecite{Li04}, the $T_1$-relaxation and
$T_2$-dephasing rates read
\bea\label{T1T2}
\frac{1}{T_1}&=&
   \frac{\sin^{2}\theta}{4} \left\{\chi^2
   \sum_{mn}{}'\left[ F(eV_{mn}+\Delta)+F(eV_{mn}-\Delta)\right] \right.  \nl
 && \left. +T_{24}\left[ F(eV_{H}+\Delta)+F(eV_{H}-\Delta)\right] \right\}, \nl
\frac{1}{T_2}&=&\frac{1}{2T_1}
    +\frac{\cos^2\theta}{2}\left[\chi^2 \sum_{mn}{}' F(eV_{mn})
    +T_{24}F(eV_H)\right]  .
\eea
Here $\chi=|\sqrt{T_{41}}-\sqrt{T_{21}}|$, and
$F(x)= x\coth(\beta x/2)$, with $\beta$ the inverse temperature.
The mixing angle ``$\theta$" is introduced by $\cos\theta=2\epsilon/\Delta$,
or $\sin\theta=2\Omega/\Delta$.
$\epsilon$ and $\Delta$ are, respectively,
the spin-up and spin-down level offset
and the eigen-energy difference, defined by
$\epsilon=(\epsilon_\uparrow-\epsilon_\downarrow)/2$,
$E_0=-\sqrt{\epsilon^2+\Omega^2}\equiv -\Delta/2$ and
$E_1=\sqrt{\epsilon^2+\Omega^2}\equiv \Delta/2$,
where $(\epsilon_\uparrow+\epsilon_\downarrow)/2$
has been taken as the reference energy (i.e. energy zero).
$V_{mn}$ is the voltage between the $m$th and $n$th reservoirs,
and the partial summation $\sum'_{mn}$ is over
transport channels ``$(m,n)=(1,2),(1,4),(2,3),(3,4)$".
In our treatment, the different back-action of electron tunneling in
channels ``$(1,3)$" and ``$(2,4)$" on the spin has been
taken into account, i.e., the former is independent of the spin state,
thus has no dissipative effect on the spin dynamics,
while the latter affects the spin significantly,
because of the opposite Hall voltages for different spin states.

Based on \Eq{T1T2} and the numerical estimate obtained above,
we find $T_1$ and $T_2$ to be about $\sim 10^{-6}$ sec.
In general, the quantum measurement time ($t_{\rm meas}$)
is in between $T_1$ and $T_2$ \cite{Sch98}.
As a consequence, if the spin coherence time, which is limited
by other scattering mechanisms, is longer than the measurement time,
the proposed Hall magnetometer would enable not only classical
measurement as already discussed above, but also quantum measurement.
To perform quantum projective measurement, this readout-time analysis
is necessary. However, to our knowledge,
similar analysis/estimate of the measurement time
based on {\it realistic device setup} is widely lacking in
the recently proposed quantum measurement schemes \cite{Gur97,Sch98,Rug0103}.

Finally, a number of practical issues/difficulties are remarked as follows:
(i) We have restricted our analysis in
the ballistic regime which makes sense for narrow Hall junctions
and/or for a high mobility 2DEG (e.g. at very low temperature).
For completeness, here we compare it with the qualitative result
in the diffusive regime which is valid under the opposite conditions.
In this case, $R_H/R_0\simeq \la B\ra \mu$, with $\mu$ the
2DEG mobility \cite{Rei00}.
Similar derivation as in the ballistic regime
leads to $V_H/(\la B\ra V_{13})=\gamma \mu$, with $\gamma$ in the order
of magnitude of unity.
For the n-type GaAs 2DEG, taking the typical value
$\mu=8.35\times 10^3 ~{\rm cm}^2/{\rm V}~{\rm sec}$
at room temperature \cite{Rei00},
we find $V_H/(\la B\ra V_{13})\simeq 0.83 ~{\rm Tesla}^{-1}$,
which is similar to the ballistic result
$V_H/(\la B\ra V_{13})=C_1/(C_2E_F)\simeq 0.96 ~{\rm Tesla}^{-1}$.
Obviously, with decreasing temperature the diffusive formula would
break down, and the Hall voltage does not linearly depend on the
increasing mobility. In this situation, the ballistic formula should be
adopted.
(ii) At finite temperature, the sensitivity of the Hall magnetometer
would be fundamentally limited by the (thermal) Johnson noise
$V_H=\sqrt{4k_BTR_sf}$, where $R_s$ and $f$ are the series resistance
of the device and the measurement bandwidth \cite{Gei03,Ora96}.
At $T=4.2$ K, the typical series resistance
$R_s=1.5 ~{\rm K}\Omega$, and the measurement bandwidth $f=1$ KHz.
We then arrive at a voltage noise about 6 nV, which
approximately reaches the detection limit (i.e. signal equals noise).
In practice, this thermal noise may be suppressed further
by decreasing temperature or improving the measurement techniques.
(iii) It would be difficult to fabricate the assumed
10 nm conducting channels. The difficulty originates from the side wall
depletion effect. Viewing the depletion layer of $\sim 100$ nm
at each side wall, controlling a width of 10 nm conducting channel
would be a challenging job.


In summary, our analysis shows that under appropriate setup design
the sub-micron Hall magnetometer can generate a Hall signal of $\sim 10$
nano-volts when measuring a single electron spin.
We thus anticipate that if the thermal-noise suppression
and fabrication difficulty can be well resolved, the
proposed Hall device is a promising candidate
of single electron spin sensor.

\vspace{5ex} {\it Acknowledgments.}
Support from the National Natural Science
Foundation of China (NNSFC), and
the Major State Basic Research Project No.\ G001CB3095 of China
are gratefully acknowledged.



\clearpage

\vspace{5ex}

\end{document}